\documentclass[a4paper]{report}
\usepackage[utf8]{inputenc}
\usepackage[T1]{fontenc}
\usepackage{RJournal}
\usepackage{amsmath,amssymb,array}
\usepackage{booktabs}

\usepackage{listings}
\begin{document}

\sectionhead{Contributed research article}
\volume{XX}
\volnumber{YY}
\year{20ZZ}
\month{AAAA}

\begin{article}

\title{A method for deriving information from running R code}
\author{by Mark P.J. van der Loo}

\maketitle

\abstract{
It is often useful to tap information from a running R script.  Obvious use
cases include monitoring the consumption of resources (time, memory) and
logging. Perhaps less obvious cases include tracking changes in R objects or
collecting output of unit tests. In this paper we demonstrate an approach that
abstracts collection and processing of such secondary information from the
running R script.  Our approach is based on a combination of three elements.
The first element is to build  a customized way to evaluate code.  The second
is labeled \emph{local masking} and it involves temporarily masking a
user-facing function so an alternative version of it is called. The third
element we label \emph{local side effect}. This refers to the fact that the
masking function exports information to the secondary information flow without
altering a global state.  The result is a method for building systems in pure R
that lets users create and control secondary flows of information with minimal
impact on their workflow, and no global side effects.
}

\section*{Introduction}
The R language provides a convenient language to read, manipulate, and write
data in the form of scripts. As with any other scripted language, an R script
gives description of data manipulation activities, one after the other, when
read from top to bottom. Alternatively we can think of an R script as a
one-dimensional visualisation of data flowing from one processing step to the
next, where intermediate variables or pipe operators carry data from one
treatment to the next. 

We run into limitations of this one-dimensional view when we want to produce
data flows that are somehow `orthogonal' to the flow of the data being treated.
For example, we may wish to follow the state of a variable while a script is
being executed, report on progress (logging), or keep track of resource
consumption. Indeed, the sequential (one-dimensional) nature of a script forces
one to introduce extra expressions between the data processing code.

As an example, consider a code fragment where the variable \code{x} is manipulated.
\begin{example}
  x[x > threshold] <- threshold
  x[is.na(x)]  <- median(x, na.rm=TRUE)
\end{example}
In the first statement every value above a certain threshold is replaced with a
fixed value, and next, missing values are replaced with the median of the
completed cases. It is interesting to know how an aggregate of interest, say
the mean of \code{x}, evolves as it gets processed. The instinctive way to do
this is to edit the code by adding statements to the script that collect the
desired information.
\begin{example}
  meanx <- mean(x, na.rm=TRUE)
  x[x > threshold] <- threshold
  meanx <- c(meanx, mean(x, na.rm=TRUE))
  x[is.na(x)]  <- median(x, na.rm=TRUE)
  meanx <- c(meanx, mean(x, na.rm=TRUE))
\end{example}
This solution clutters the script by insterting expresssions that are not
necessary for its main purpose. Moreover, the tracking statements are
repetitive, which validates some form of abstraction.

\begin{figure}[t]
\centering
\includegraphics[width=\textwidth]{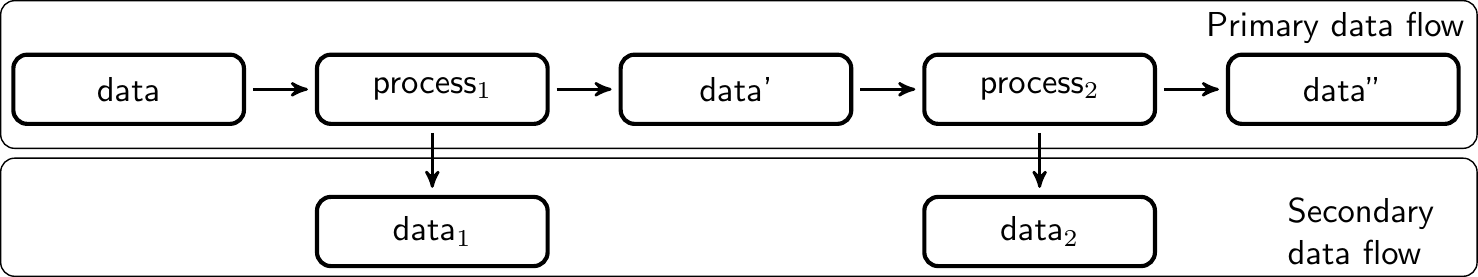}
\caption{Primary and secondary data flows in an R script. The primary flow
follows the execution of an R script, while in the background a secondary data
flow (e.g.\ logging information) is created.}
\label{fig:streams}
\end{figure}

A more general picture of what we would like to achieve is given in
Figure~\ref{fig:streams}. The `primary data flow' is developed by a user as a
script. In the previous example this concerns processing \code{x}.  When the
script runs, some kind of logging information, which we label the `secondary
data flow' is derived implicitly by an abstraction layer. 

Creating an abstraction layer means that concerns between primary and secondary
data flows are separated as much as possible. In particular, we want to prevent
the abstraction layer from inspecting or altering the user code that describes
the primary data flow. Furthermore, we would like the user to have some control
over the secondary flow from within the script, for example to start, stop or
parameterize the secondary flow. This should be done with minumum editing of
the original user code and it should not rely on global side effects. This
means that neither the user, nor the abstraction layer for the secondary data
flow should have to manipulate or read global variables, options, or other
environmental settings to convey information from one flow to the other.
Finally, we want to treat the availability of a secondary data flow as a normal
situation. This means we wish to avoid using signaling conditions (e.g.\
warnings or errors) to convey information between the flows, unless there is an
actual exceptional condition such as an error.

\subsection{Prior art}
There are several packages that generate a secondary data flow from a running
script. One straightforward application concerns logging messages that report
on the status of a running script. To create a logging message, users edit
their code by inserting logging expressions where desired. Logging expressions
are functions calls that help building expressions, for example by
automatically adding a time stamp. Configuration options usually include a
measure of logging verbosity, and setting an output channel that controls where
logging data will be sent.  Changing these settings relies on communication
from the main script to the functionality that controlls the flow of logging
data.  In \pkg{logger} \citep{daroczi2019logger} this is done by manipulating a
variable stored in the package namespace using special helper functions.  The
\pkg{logging} package \citep{frasca2019logging} also uses an environment within
the namespace of the package to manage option settings, while
\pkg{futile.logger} \citep{rowe2016futile.logger} implements a custom global
option settings manager that is somewhat comparable to R's own \code{options()}
function. 

Packages \pkg{bench} \citep{hester2019bench} and \pkg{microbenchmark}
\citep{mersmann2018microbenchmark} provide time profiling of single R
expressions. The \pkg{bench} package also includes memory profiling.  Their
purpose is not to derive a secondary data flow from a running production script
as in Figure~\ref{fig:streams} but to compare performance of R expressions.
Both packages export a function that accepts a sequence of expressions to
profile.  These functions take control of expression execution and insert time
and/or memory measurements where necessary. Options, such as the number of
times each expression is executed, are passed directly to the respective
function. 

Unit testing frameworks provide another source of secondary data flows.  Here,
an R script is used to prepare, setup, and compare test data, while the results
of comparisons are tapped and reported. Testing frameworks are provided by
\pkg{testthat} \citep{wickham2011testthat}, \pkg{RUnit},
\citep{burger2018runit}, \pkg{testit} \cite{xie2018testit}, \pkg{unitizer}
\citep{gaslam2019unitizer}, and \pkg{tinytest} \citep{loo2019tinytest}. The
first three packages (\pkg{testthat}, \pkg{RUnit} and \pkg{testit}) all export
assertion functions that generate condition signals to convey information about
test results. Packages \pkg{RUnit} and \pkg{testit} use \code{sys.source()} to
run a file containing unit test assertions and exit on first error while
\pkg{testthat} uses \code{eval()} to run expressions, capture conditions and
test results and reports afterwards. The \pkg{unitizer} framework is different
because it implements an interactive prompt to run tests and explore their
results. Rather than providing explicit assertions, \pkg{unitizer} stores
results of all expressions that return a visible result and compares their
output at subsequent runs.  Interestingly, \pkg{unitizer} allows for optional
monitoring of the testing environment. This includes environment variables,
options, and more. This is done by manipulating code of (base) R functions that
manage these settings and masking the original functions temporarily. These
masking functions then provide parts of the secondary data flow (changes in the
environment).  Finally, \pkg{tinytest} is based on the approach that is the
topic of this paper and it will be discused as an application below.

Finally we note the \pkg{covr} package of \citet{hester2018covr}. This package
is used to keep track of which expressions of an R package are run (covered) by
package tests or examples. In this case the primary data flow is a test
script executing code (functions, methods) stored in another script, usually in
the context of a package. The secondary flow consists of counts of how often
each expression in the source files are executed. The package works by parsing
and altering the code in the source file, inserting expressions that increase
appropriate counters. These counters are stored in a variable that is part of
the package's namespace.

Summarizing, we find that in logging packages the secondary data flow is
invoked explicitly by users while configuration settings are communicated by
manipulating a global state that may or may not be directly accessible by the
user. For benchmarking packages, the expressions are passed explicitly to an
`expression runner' that monitors effect on memory and passage of time.  In
most test packages the secondary flow is invoked explicitly using special
assertions that throw condition signals. Test files are run using functionality
that captures and administrates signals where necessary. Two of the discussed
packages explicitly manipulate existing code before running it to create a
secondary data flow. The \pkg{covr} package does this to update expression
counters and the \pkg{unitizer} package to monitor changes in the global state.

\subsection{Contribution of this paper}
The purpose of this paper is to first provide some insight into the problem of
managing multiple data flows, independent of specific applications. In the
following section we discuss managing a secondary data stream from the point of
view of changing the way in which expressions are combined and executed by R.

Next, we highlight two programming patterns that allow one to derive a
secondary data stream, both in non-interactive (while executing a file) and in
interactive circumstances. The methods discussed here do not require explicit
inspection or modification of the code that describes the primary data flow. It
is also not necessary to invoke signalling conditions to transport information
from or to the secondary data stream.

We also demonstrate a combination of techniques that allow users to
parameterize the secondary flow, without resorting to global variables, global
options, or variables within a package's namespace. We call this technique
`local masking' with `local side effects'. It is based on temporarily and
locally masking a user-facing function with a function that does exactly the
same except for a side effect that passes information to the secondary data
flow. 

As examples we discuss two applications where these techniques have been
implemented. The first is the \pkg{lumberjack} package
\citep{loo2019lumberjack}, which allows for tracking changes in R objects as
they are manipulated expression by expression. The second is \pkg{tinytest}
\citep{loo2019tinytest}, a compact and extensible unit testing framework.

Finally, we discuss some advantages and limitations to the techniques proposed.

\section{Concepts}
In this section we give a high-level overview of the problem of adding a second
data flow to an existing one, as well as a general way to think about a
solution. The general approach was inspired by a discussion of
\citet{milewski2018category} and is related to what is sometimes
called a \emph{bind operator} in functional programming. 

Consider as an example the following two expressions, labeled $e_1$ and $e_2$.
\begin{example}
  e1:  x <- 10 
  e2:  y <- 2*x
\end{example}
We would like to implement some kind of monitoring as these expressions are
evaluated. For this purpose it is useful to think of think of $e_1$ and $e_2$
as functions that accept a set of key-value pairs, possibly alter the set's
contents, and return it. In R this set of key-value pairs is an
\code{environment}, and usually it is the global environment (the user's
workspace). Starting with an empty environment $\{\}$ we get:
\begin{align*}
 e_1(\{\})        &= \{(\texttt{"x"},\texttt{10})\}\\
 e_2(e_1(\{\})) &= \{(\texttt{"x"}, \texttt{10}),(\texttt{"y"}, \texttt{20})\}
\end{align*}
In this representation we can write the result of executing the above script in
terms of the function composition operator $\circ$:
\begin{align*}
e_2(e_1(\{\})) = (e_2\circ e_1)(\{\}).
\end{align*}
And in general we can express the final state $\mathcal{U}$ of any environment after
executing a sequence of expressions $e_1,e_2,\cdots,e_k$ as
\begin{align}
\mathcal{U} = (e_k\circ e_{k-1}\circ \cdots\circ e_1)(\{\}),
\label{eq:compose}
\end{align}
where we assumed without loss of generality that we start with an empty
environment. We will refer to the sequence $e_1\ldots e_k$ as the `primary
expressions', since they define a user's main data flow. 

We now wish to introduce some kind of logging. For example, we want to count
the number of evaluated expressions, not counting the expressions that will
perform the count. The naive way to do this is to introduce a new expression,
say $n$:
\begin{example}
  n:   if (!exists("N")) N <- 1 else N <- N + 1
\end{example}
And we insert this into the original sequence of expressions. This amounts to the
cumbersome solution
\begin{align}
\mathcal{U}\cup\{(\texttt{"N"},k)\} = (n\circ e_k \circ n\circ e_{k-1}\circ n\circ \cdots n\circ e_{1})(\{\}),
\label{eq:insert}
\end{align}
where the number of executed expressions is stored in \code{N}. We shall refer
to \code{n} as a `secondary expression' as it does not contribute to the user's
primary data flow.

The above procedure can be simplified if we define a new function composition
operator $\circ_n$ as follows.
\begin{align*}
a\circ_n b = a\circ n \circ b.
\end{align*}
One may verify  the associativity property $a\circ_n(b\circ_n c)=(a\circ_n
b)\circ_n c$ for expressions $a$, $b$ and $c$, so $\circ_n$ can indeed be
interpreted as a new function composition operator. Using this operator we get
\begin{align}
\mathcal{U}\cup\{(\texttt{"N"},k-1)\} = (e_k\circ_n e_{k-1}\circ_n \cdots\circ_n e_1)(\{\}),
\label{eq:compose}
\end{align}
which gives the same result as Equation~\ref{eq:insert} up to a constant.

If we are able to alter function composition, then this mechanism can be used
to track all sorts of useful information during the execution of $e_1,\ldots,
e_k$.  For example, a simple profiler is set up by timing the expressions and
adding the following expression to the function composition operator.
\begin{example}
  s: if (!exists("S")) S <- Sys.time() else S <- c(S, Sys.time())
\end{example} 
After running $e_k\circ_s\cdots\circ_s e_1$,  \code{diff(S)} gives the timings of
individual statements. A simple memory profiler is defined as follows.
\begin{example}
  m: if (!exists("M")) M <- sum(memory.profile()) else M <- c(M, sum(memory.profile()))
\end{example}
After running  $e_k\circ_m\cdots\circ_m e_1$, \code{M} gives the amount of memory
used by R after each expression.

We can also track changes in data, but it requires that the composition
operator knows the name of the R object that is being tracked.
As an example, consider the following primary expressions.
\begin{example}
  e1:  x <- rnorm(10)
  e2:  x[x<0] <- 0
  e3:  print(x)
\end{example}
We can define the following expression for our modified function composition
operator.
\begin{example}
  v:  {
        if (!exists("V")){ 
          V <- logical(0)
          x0 <- x
        }
        if (identical(x0,x)) V <- c(V, FALSE)
        else V <- c(V, TRUE)
        x0 <- x
      }
\end{example}
After running $e_3\circ_v e_2\circ_v e_1$ the variable \code{V} equals
\code{c(TRUE, FALSE)}, indicating that $e_2$ changed \code{x} and $e_3$ did
not.

These examples demonstrate that redefining function composition yields a
powerful method for extracting logging information with (almost) no intrusion
on the user's common work flow. The simple model shown here does have some
obvious setbacks: first, the expressions inserted by the composition operator
manipulate the same environment as the user expressions. The user- and
secondary expressions can therefore interfere with each other's results.
Second, there is no direct control from the primary sequence over the secondary
sequence: the user has no explicit control over starting, stopping, or
parametrizing the secondary data stream.  We demonstrate in the next section
how these setbacks can be avoided by evaluating secondary expressions in a
separate environment, and by using a techniques we call `local masking' and
`local side-effects'.

\section{Creating a secondary data flow with R}
R executes expressions one by one in a read-evaluate-print loop (REPL). In
order to tap information from this running loop it is necessary to catch the
user's expressions and interweave them with our own expressions. One way to do
this is to develop an alternative to R's native \code{source()} function.
Recall that \code{source()} reads an R script and executes all expressions in
the global environment. Applications include non-interactive sessions or
interactive sessions with repetitive tasks such as running test scripts while
developing functions. A second way to intervene with a user's code is to
develop a special `forward pipe' operator, akin to for example the
\pkg{magrittr} pipe of \citet{bache2014magrittr} or the `dot-pipe' of
\citet{mount2018dot}. Since a user inserts a pipe between expressions, it is an
obvious place to insert code that generates a secondary data flow.

In the following two subsections we will develop both approaches. As a running
example we will implement a secondary data stream that counts expressions. 

\subsection{Build your own \code{source()}}
The \code{source()} function reads an R script and executes all expressions in
the global environment. A simple variant of \code{source()} that counts
expressions as they get evaluated can be built using \code{parse()} and
\code{eval()}.
\begin{example}
  run <- function(file){
    expressions <- parse(file)
    runtime <- new.env(parent=.GlobalEnv)

    n <- 0
    for (e in expressions){ 
      eval(e, envir=runtime)
      n <- n + 1
    }
    message(sprintf("Counted 
    runtime
  }
\end{example}
Here \code{parse()} reads the R file and returns a list of expressions
(technically, an object of class `\code{expression}').  The \code{eval()} function
executes the expression while all variables created by, or needed for
execution are sought in a newly created environment called \code{runtime}.  We
make sure that variables and functions in the global environment are found by
setting the parent of \code{runtime} equal to \code{.GlobalEnv}.  Now, given a
file \code{"script.R"}.
\begin{example}
  # contents of script.R
  x <- 10
  y <- 2*x
\end{example}
An interactive session would look like this.
\begin{example}
  > e <- run("script.R")
  Counted 2 expressions
  > e$x
  [1] 10
\end{example}
So contrary to default behavior of \code{source()}, variables are assigned in a
new environment. This difference in behavior can be avoided by evaluating
expressions in \code{.GlobalEnv}, but for the next step it is important to have
a separate runtime environment.

We now wish to give the user some control over the secondary data stream.  In
particular, we want the user to be able to choose when \code{run()} starts
counting expressions. Recall that we demand that this is done by direct
communication to \code{run()}. This means that side-effects such as setting a
special variable in the global environment or a a global option is out of the
question. Furthermore, we want to avoid code inspection: the \code{run()}
function should be unaware of what expressions it is running exactly.  We start
by writing a function for the user that returns \code{TRUE}.
\begin{example}
  start_counting <- function() TRUE
\end{example}
Our task is to capture this output from \code{run()} when \code{start\_counting()}
is called. We do this by masking this function  with another function that does
exactly the same, except that it also copies the output value to a place where
\code{run()} can find it. To achieve this, we use the following helper function.
\begin{example}
  capture <- function(fun, envir){
    function(...){
      out <- fun(...)
      envir$counting <- out
      out
    }
  }
\end{example}
This function accepts a function (\code{fun}) and an environment
(\code{envir}). It returns a function that first executes \code{fun(...)},
copies its output value to \code{envir} and then returns the output to the
user. In an interactive session, we would see the following.
\begin{example}
  > store <- new.env()
  > f <- capture(start_counting, store)
  > f()
  [1] TRUE
  > store$counting
  [1] TRUE
\end{example}
Observe that our call to \code{f()} returns \code{TRUE} as expected, but also
exported a copy of \code{TRUE} into \code{store}.  The reason this works is
that an R function `remembers' where it is created.  The function \code{f()}
was created inside \code{capture()} and the variable \code{envir} is present
there. We say that this `capturing' version of \code{start\_counting} has a
\emph{local side-effect}: it writes outside of its own scope but the place
where it writes is controlled.

We now need to make sure that \code{run()} executes the captured version of
\code{start\_counting()}.  This is done by locally masking the user-facing
version of \code{start\_counting()}. That is, we make sure that the captured
version is found by \code{eval()} and not the original version. A new version of
\code{run()} now looks as follows.
\begin{example}
  run <- function(file){
    expressions <- parse(file)
    store <- new.env()
    runtime <- new.env(parent=.GlobalEnv)
    runtime$start_counting <- capture(start_counting, store)
    n <- 0
    for (e in expressions){ 
      eval(e, envir=runtime)
      if ( isTRUE(store$counting) ) n <- n + 1
    }
    message(sprintf("Counted 
    runtime
  }
\end{example}
Now, consider the following code, stored in \code{script1.R}.
\begin{example}
  # contents of script1.R
  x <- 10
  start_counting()
  y <- 2*x
\end{example}
In an interactive session we would see this.
\begin{example}
  > e <- run("script1.R")
  Counted 1 expressions
  > e$x
  [1] 10
  > e$y
  [1] 20
\end{example}

Let us go through the most important parts of the new \code{run()} function.
After parsing the R file  a new environment is created that will store the
output of calls to \code{start\_counting()}.
\begin{example}
    store  <- new.env()
\end{example}
The runtime environment is created as before, but now we add the capturing
version of \code{start\_counting()}.
\begin{example}
    runtime <- new.env(parent=.GlobalEnv)
    runtime$start_counting <- capture(start_counting, store)
\end{example}
This ensures that when the user calls \code{start\_counting()}, the capturing
version is executed. We call this technique \emph{local masking} since the
\code{start\_counting()} function is only masked during the execution of
\code{run()}.  
The captured version of \code{start\_counting()}as a side effect stores its
output in \code{store}.  We call this a `local side-effect' because
\code{store} is never seen by the user: it is created inside \code{run()} and
destroyed when \code{run()} is finished.

Finally, all expressions are executed in the runtime environment and counted
conditional on the value of
\code{store\$counting}. 
\begin{example}
    for (e in expressions){ 
      eval(e, envir=runtime)
      if ( isTRUE(store$counting) ) n <- n + 1
    }
\end{example}

Summarizing, with this construction we are able to create a file runner akin to
\code{source()} that can gather and communicate useful process metadata while
executing a script. Moreover, the user of the script can convey information
directly to the file runner, while it runs, without relying on global
side-effects. This is achieved by first creating a user-facing function that
returns the information to be send to the file runner. The file runner locally
masks the user-facing version with a version that copies the output to an
environment local to the file runner before returning the output to the user. 

The approach just described can be generalized to more realistic use cases. All
examples mentioned in the `Context' section ---time or memory profiling, or
logging changes in data, merely need some extra administration.  Furthermore,
the current example emits the secondary data flow as a `\code{message}'.  In
practical use cases it may make more sense to write the output to a file
connection or database, or the make the secondary data stream output of the
file runner. In the Applications section both applications are discussed.

\subsection{Build your own pipe operator}
The \pkg{magrittr} forward `pipe' operator of \citet{bache2014magrittr} has
become a popular tool for R users over the last years. This pipe operator is
intended as a form of `syntactic sugar' that in some cases makes code a little
easier to write. A pipe operator behaves somewhat like a left-to-right
`expression composition operator'. This, in the sense that a sequence of
expressions that are joined by a pipe operator are interpreted by R's parser as
a single expression. Pipe operators also offer an opportunity to derive
information from a running sequence of expressions.

The \pkg{magrittr} pipe operator has quite complex semantics, but it is
possible to implement a basic pipe operator as follows.
\begin{example}
  `
\end{example}
Here, the \code{rhs} (right hand side) argument must be a single-argument
function, which is applied to \code{lhs}. In an interactive session we could
see this.
\begin{example}
  > 3 
  [1] 0.9900591
\end{example}

To build our expression counter, we need to have a place to store the counter
value, hidden from the user. In contrast to the implementation of the file
runner in the previous section, each use of \code{\%p>\%} is disconnected from
the other, and there seems to be no shared space to increase the counter at
each call. The solution is to let the secondary data flow travel with the
primary flow, by adding an attribute to the data. We create two user-facing
functions that start or stop logging, as follows.
\begin{example}
  start_counting <- function(data){
    attr(data, "n") <- 0
    data
  }
  end_counting <- function(data){
    message(sprintf("Counted 
    attr(data, "n") <- NULL
    data
  }
\end{example}
Here the first function attaches a counter to the data and initializes it to
zero. The second function reports its value, decreased by one so the stop
function itself is not included in the count.  We also alter the pipe operator
to increase the counter, if it exists.
\begin{example}
  `
    if ( !is.null(attr(lhs,"n")) ){
      attr(lhs,"n") <- attr(lhs,"n") + 1
    }
    rhs(lhs)
  }
\end{example}
In an interactive session, we could now see the following.
\begin{example}
  > out <- 3 
  +  start_counting 
  +    sin 
  +    cos 
  +  end_counting
  Counted 2 expressions
  > out
  [1] 0.9900591
\end{example}

Summarizing, for small interactive tasks a secondary data flow can be added to
the primary one by using a special kind of pipe operator. Communication between
the user and the secondary data flow is implemented by adding or altering
attributes attached to the R object.

Generalizations of this technique come with a few caveats. First, the current
pipe operator only allows right-hand side expressions that accept a single
argument. Extension to a more general case involves inspection and manipulation
of the right-hand side's abstract syntax tree and is out of scope for the
current work. Second, the current implementation relies on the right-hand side
expressions to preserve attributes. A general implementation will have to test
that the output of rhs(lhs) still has the logging attribute attached (if there
was any) and re-attach it if necessary.

\section{Application 1: tracking changes in data}
\label{sect:examples}

The \pkg{lumberjack} package \citep{loo2019lumberjack} implements a logging
framework to track changes in R objects as they get processed. The package
implements both a pipe operator, denoted \code{\%L>\%} and a file runner called
\code{run\_file()}. The main communication devices for the user are two
functions called \code{start\_log()} and \code{dump\_log()}. 

We will first demonstrate working with the \pkg{lumberjack} pipe operator.
The function \code{start\_log()} accepts an R object and a logger object. It
attaches the logger to the R object and returns the augmented R object. A
logger is a reference object\footnote{A native R Reference Class, an
`\code{R6}' object \citep{chang2019r6} or any other reference type object
implementing the proper API.} that exposes at least an \code{\$add()} method
and a \code{\$dump()} method.  If a logger is present, the pipe operator stores
a copy of the left hand side.  Next, it executes the expression on the
right-hand side with the left-hand side as an argument and stores the output.
It then calls the \code{add()} method of the logger with the input and output,
so that the logger can compute and store the difference.  The
\code{dump\_log()} function accepts an R object, calls the \code{\$dump()}
method on the attached logger (if there is any), removes the logger from the
object and returns the object. An interactive session could look as follows.
\begin{example}
  > library(lumberjack)
  > out <- women 
  >   start_log(simple$new()) 
  >   transform(height = height * 2.54) 
  >   identity() 
  >   dump_log()
  Dumped a log at /home/mark/simple.csv
  > read.csv("simple.csv")
    step                time                        expression changed
  1    1 2019-08-09 11:29:06 transform(height = height * 2.54)    TRUE
  2    2 2019-08-09 11:29:06                        identity()   FALSE
\end{example}
Here, \code{simple\$new()} creates a logger object that registers whether an R
object has changed or not. There are other loggers that compute more involved
differences between in- and output. The \code{\$dump()} method of the logger
writes the logging output to a csv file.

For larger scripts, a file runner called \code{run\_file()} is available in
\pkg{lumberjack}. As an example consider the following script. It converts
columns of the built-in \code{women} data set to SI units (meters and kilogram)
and then computes the body-mass index of each case.
\begin{example}
  # contents of script2.R
  start_log(women, simple$new())
  women$height <- women$height * 2.54/100
  women$weight <- women$weight * 0.453592
  women$bmi    <- women$weight/(women$height)^2
\end{example}
In an interactive session we can run the script and access both the logging
information and retrieve the output of the script.
\begin{example}
  > e <- run_file("script2.R")
  Dumped a log at /home/mark/women_simple.csv
  > read.csv("women_simple.csv")
    step                time                                 expression changed
  1    1 2019-08-09 13:11:25             start_log(women, simple$new())   FALSE
  2    2 2019-08-09 13:11:25    women$height <- women$height * 2.54/100    TRUE
  3    3 2019-08-09 13:11:25    women$weight <- women$weight * 0.453592    TRUE
  4    4 2019-08-09 13:11:25 women$bmi <- women$weight/(women$height)^2    TRUE
  > head(e$women,3)
    height   weight      bmi
  1 1.4732 52.16308 24.03476
  2 1.4986 53.07026 23.63087
  3 1.5240 54.43104 23.43563
\end{example}

The \pkg{lumberjack} file runner locally masks  \code{start\_log()} with a
function that stores the logger and the name of the tracked R object in a local
environment. A copy of the tracked object is stored locally as well.
Expressions in the script are executed one by one. After each expression, the
object in the runtime environment is compared with the stored object.  If it
has changed, the \code{\$add()} method of the logger is called and a copy of the
changed object is stored. After all expressions have been executed, the
\code{\$dump()} method is called so the user does not have to do this explicitly.

A user can add multiple loggers for each R object and track multiple objects.
It is also possible to dump specific logs for specific objects during the
script.  All communication necessary for these operations runs via the
mechanism explained in the `build your own \code{source()}' section.

\section{Application 2: unit testing}
The \pkg{tinytest} package \citep{loo2019tinytest} implements a unit testing
framework. Its core function is a file runner that uses local masking and local
side effects to capture the output of assertions that are inserted explictly by
the user.  As an example, we create tests for the following function. 
\begin{example}
  # contents of bmi.R
  bmi <- function(weight, height) weight/(height^2)
\end{example}
A simple \pkg{tinytest} test file could look like this.
\begin{example}
  # contents of test_script.R
  data(women)
  women$height <- women$height * 2.54/100
  women$weight <- women$weight * 0.453592
  BMI    <- with(women, bmi(weight,height) )

  expect_true( all(BMI >= 10) )
  expect_true( all(BMI <= 30) )
\end{example}
The first four lines prepare some data, while the last two lines check whether
the prepared data meets our expectations.  In an interactive session we can run
the test file, after loading the \code{bmi()} function.
\begin{example}
  > source("bmi.R")
  > library(tinytest)
  > out <- run_test_file('test_script.R')
  Running test_script.R.................    2 tests OK
  > print(out, passes=TRUE)
  ----- PASSED      : test_script.R<7--7>
   call| expect_true(all(BMI >= 10))
  ----- PASSED      : test_script.R<8--8>
   call| expect_true(all(BMI <= 30))
\end{example}

In this application, the file runner locally masks the \code{expect\_*()} functions
and captures their result through a local side effect. As we are only
interested in the test results, the output of all other expressions is
discarded.

Compared to the basic version described in the `build your own \code{source()}'
section, this file runner keeps some extra administration such as the line
numbers of each masked expression. These can be extracted from the output of
\code{parse()}. The package comes with a number of assertions in the form of
\code{expect\_*()} functions. It is possible to extend \pkg{tinytest} by
registering new assertions. These are then automatically masked by the file
runner. The only requirement on the new assertions is that they return an
object of the same type as the built-in assertions (an object of class
`\code{tinytest}').

\section{Discussion}
\label{sect:discussion} 
The techniques demonstrated here have two major advantages. First, it allows
for a clean and side-effect free separation between the primary and secondary
data flows. As a result, the secondary data flow is composes with the primary
data flow. In other words: a user that wants to add a secondary data flow to an
existing script does not have to edit any existing code. Instead it is only
necessary to add a bit of code to specify and initialize the secondary stream,
which is a big advantage for maintainability.  Second, the current mechanisms
avoid the use of condition signals. This also leads to code that is easier to
understand and navigate because all code associated with the secondary flow can
be limited to the scope of a single function (here: either a file runner or a
pipe operator). Since the secondary data flow is not treated as an unusual
condition (exception) the exception signaling channel is free for transmitting
truly unusual conditions such as errors and warnings.

There are also some limitations inherent to these techniques. Although the code
for the secondary data flow is easy to compose with code for the primary data
flow, it is not as easy to compose different secondary data flows. For example:
one can use only one file runner to run an R script, and only a single pipe
operator to combine two expressions. 

A second limitation is that this approach does not recurse into the primary
expressions. For example, the expression counters we developed only count
user-defined expressions: they can not count expressions that are called by
functions called by the user. This means that something like a code coverage
tool such as \pkg{covr} is out of scope.

A third and related limitation is that the resolution of expressions may be too
low for certain applications. For example in R, `\code{if}' is an expression
(it returns a value when evaluated) rather then a statement (like \code{for}).
This means that \code{parse()} interprets a block such as
\begin{example}
  if ( x > 0 ){
   x <- 10
   y <- 2*x
  }
\end{example}
as a single expression. If higher resolution is needed, this requires explicit
manipulation of the user code. 

Finally, the local masking mechanism excudes the use of the namespace
resolution operator. For example, in \pkg{lumberjack} it is not possible to use
\code{lumberjack::start\_log()} since in that case the user-facing function
from the package is executed and not the masked function with the desired local
side-effect.

\section{Conclusion}
\label{sect:conclusion}
In this paper we demonstrated a set of techniques that allow one to add a
secondary data flow to an existing user-defined R script. The core idea is that
we manipulate way expressions are combined before they are executed. In
practice, we use R's \code{parse()} and \code{eval()} to add secondary data
stream to user code, or build a special `pipe' operator. Local masking and
local side effects allow a user to control the secondary data flow without
global side-effects. The result is a clean separation of concerns between the
primary and secondary data flow, that does not rely on condition handling,
is void of global side-effects, and that is implemented in pure R.

\address{Mark P.J. van der Loo\\
Statistics Netherlands\\
PO-BOX 24500, 2490HA Den Haag\\
The Netherlands\\
\href{https://orcid.org/0000-0002-9807-4686}{\includegraphics[height=\fontcharht\font`\0]{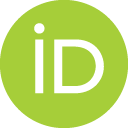}} \url{https://orcid.org/0000-0002-9807-4686}\\
\url{https://www.markvanderloo.eu}\\
\email{m.vanderloo@cbs.nl}\\
}

\bibliography{vanderloo}

\end{article}

\end{document}